# Switchable domains in point contacts based on transition metal tellurides

*Yurii G. Naidyuk[1]\*, Dima L. Bashlakov[1], Oksana E. Kvitnitskaya[1], Boy. R. Piening[2], Grigory Shipunov[2], Dmitri V. Efremov[2]\*, Saicharan Aswartham[2]\*, Bernd Büchner [2,3]*

[1]B. Verkin Institute for Low Temperature Physics and Engineering, NAS of Ukraine, 61103 Kharkiv, Ukraine

[2]Leibniz Institute for Solid State and Materials Research Dresden, Helmholtzstraße 20, D-01069 Dresden, Germany

[3]Institut für Festkörper- und Materialphysik, Technische Universität Dresden, D-01069 Dresden, Germany

**Abstract:**

We report resistive switching in voltage biased point contacts (PCs) based on series of van der Waals transition metals tellurides (TMTs) such as *Me*Te$_2$ (*Me*=Mo, W) and Ta*Me*Te$_4$ (*Me*= Ru, Rh, Ir). The switching occurs between a low resistive "metallic-type" state, which is the ground state, and a high resistive "semiconducting-type" state by applying certain bias voltage (<1V), while reverse switching takes place by applying voltage of opposite polarity. The origin of the effect can be formation of domain in PC core by applying a bias voltage, when a strong electric field (about 10kV/cm) modifies the crystal structure and controls its polarization. In addition to the discovery of the switching effect in PCs, we also suggest a simple method of material testing before functionalizing them, which offers a great advantage in finding suitable novel substances. The new functionality of studied TMTs arising from switchable domains in submicron hetero-structures that are promising, e.g., for non-volatile resistive random access memory (RRAM) engineering.

\* Corresponding authors:

naidyuk@ilt.kharkov.ua

d.efremov@ifw-dresden.de

s.aswartham@ifw-dresden.de

**Introduction**

Nowadays layered van der Waals transition metal chalcogenides (TMC) attract a lot of attention in the scientific community both due to the manifestation of their remarkable electronic properties, which can be modified by element composition, crystal structure, thickness, electronic density etc. and potential for nano-electronic and spintronic applications (see [1] and Refs. therein). It was also found that some of them (e.g., $MoTe_2$, $WTe_2$) are Weyl semimetals possessing of massless fermionic excitations and topologically robust electronic surface states [2]. On the other hand, layered structure of TMC with a weak van der Waals binding between layers allows exfoliate these materials up to monolayers. The latter opens an amazing window to observe new size dependent properties in TMC and allows take advantage of this opportunity in future applications [3].

Here we study transition metal telluride (TMT) $MoTe_2$ together with its sister compounds $WTe_2$, Ta*Me*$Te_4$ (*Me* =Ir, Rh, Ru), which have recently drawn enormous attention due to their unique physical properties, such as extremely large magnetoresistance [4-6], superconductivity [7-9], higher-order topology [10-12], and the polar metal [13-17]. $MoTe_2$ can be grown in two modifications: semiconductor 2H, and semimetal $T_d$ (at T< 250K) or T` (T>250K). The authors of Refs. [13, 14] have shown that under a high electric field the semiconductor 2H phase of $MoTe_2$ or $Mo_{1-x}W_xTe_2$ (x<0.1) (tens of nm thick), sandwiched between metallic electrodes, undergoing a transition to a transient metallic-type $2H_d$ phase. Thus, applying high electric field, the $MoTe_2$ changes resistivity from the high resistive state to the low–resistive state and vice versa *(NB. They observed also LRS to HRS switching)*.

In our study we apply well-established point contact (PC) technique [18], which allows study properties of materials in restricted geometry from submicron to nanometer scale under high current densities and electric fields. A phase transition in the sample can also be detected by measuring a nonlinear conductivity of PCs, as it was shown, for example, for ferromagnetic metals [19]. Furthermore, the method was recently successfully applied to TMT for investigations of the interface superconductivity and electron-phonon coupling constants [8, 9]. In particular, we found that at the PC the critical superconducting temperature in $MoTe_2$ is fifty times enhanced in comparison to the bulk system [8]. Interestingly, that the same strong increase of the superconducting critical temperature in $MoTe_2$ was found in monolayers [20].

In this new study of TMT, we report our finding of bi-stable resistive states with reversible switching in current-voltage *I–V* characteristics of $MoTe_2$ PCs at biases of several hundreds of millivolts. In contrast to the aforementioned works [13, 14], we start with the semimetallic ground state ($T_d$ or T`) and transfer the system to the semiconducting-like. The similar effect has been registered also for PCs based on related TMTs like $WTe_2$, $TaRuTe_4$, $TaRhTe_4$ and $TaIrTe_4$. The observed bipolar resistive switching in devices based on studied series of TMT compounds, regardless of its inner nature, is of great importance and could be applicable to create a new type of resistive random-access memory (RRAM) devices as potential alternatives to existing nonvolatile memories. Additionally, along with this discovery, the understanding underlying physics of switching processes is of particular interest, while utilizing of topological properties and surface states at interface in TMTs can pave the way to nano-electronics with astonishing functionalities.

**Results**

Figure 1 shows *dV/dI* and *I–V* characteristics of "soft" PCs (see Experimental section) made by putting of small drop of silver paint on the surface of TaIrTe$_4$ and MoTe$_2$ PCs, respectively (see Methods for more details). A closed loop for *dV/dI* and a "butterfly" shape for *I–V* curves are seen by scanning bias voltage from one to opposite polarity and back. For example, starting from the low resistance state (LRS), switching to the high resistance state (HRS) occurs above +250 mV (Fig. 1(a), black curve), while by scanning LRS to the opposite negative bias no switching is observed. At the same time, starting from the HRS (Fig. 1(a), red curve), where *dV/dI* demonstrate sharp zero-bias maximum, transition to the LRS takes place only at negative polarity, a bit above -300mV, while at positive polarity no switching to LRS is observed up to +520 mV.

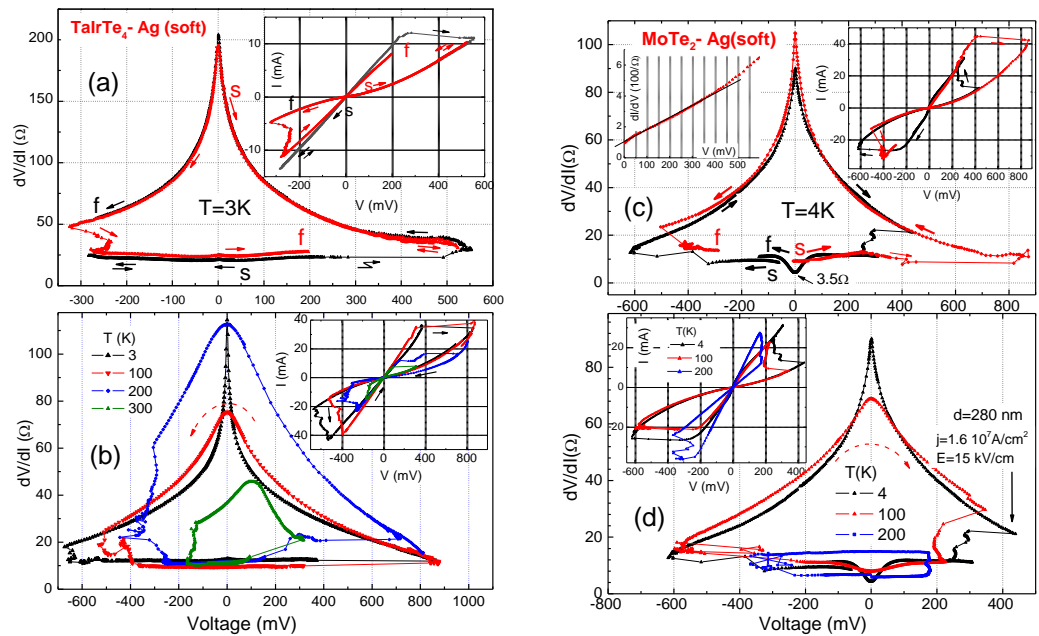

**Figure 1.** Differential resistance *dV/dI* and *I-V* curves behavior for "soft" contacts based on TMT at different temperatures. (a) *dV/dI* and *I–V* curves (inset) of "soft" contacts TaIrTe$_4$–Ag measured at 3K starting from LRS (black curve) and HRS (red curve)."s" marks the start of measurements and "f" is for the finish, arrows show direction of the current scan. The black curve is measured starting towards negative voltages - forth and back, while transition to HRS occurs only at positive voltages. The red curve is measured starting towards positive voltages - forth and back, while transition scan to LRS occurs only at negative voltages. (b) *dV/dI* and *I–V* curves (inset) of another "soft" contact TaIrTe$_4$–Ag measured at different temperatures. The curves are measured "counter clock-wise". (c) *dV/dI* of two (black and red) "soft" contacts MoTe$_2$–Ag measured starting from LRS. "s" marks the start of measurements and "f" is for the finish. Arrows show direction of the current scan. Transition from the LRS to HRS (or from the HRS to LRS) occurs at opposite polarities for these two contacts. Right inset shows *I–V* curves corresponding to *dV/dI* from the main panel. Left inset shows that the differential conductivity *dI/dV* for contact from the main panel. (d) *dV/dI* and *I–V* curves (inset) of 3.5 Ω "soft" contact from the panel (c) measured at different temperatures.

Figure 1(b) shows *dV/dI* and *I–V* characteristics of another "soft" PC based on TaIrTe$_4$ at different temperatures. Here, the switching effect persists up to the room temperature, only amplitude of zero bias *dV/dI* maximum goes down and switching voltage decreases.

Figure 1(c) demonstrates similar dependences for two MoTe$_2$ PCs. Here, the switching LRS to HRS is at positive polarity for a red curve, like for PC from Fig. 1(a), while for another PC (black curve) the switching LRS to HRS (or HRS to LRS) is at negative (positive) polarity. Interesting, that *dI/dV=(dV/dI)$^{-1}$* (left inset) displays a clear linear relationship at 4K.

Figure 1(d) shows evolution of *dV/dI* and *I–V* (in inset) characteristics of the latter PC with temperature, where amplitude of the effect and switching voltage decreases. The diameter of this PC estimated as 280 nm using temperature dependence of their zero bias resistance analogously as it was made in Supplement of Ref. [9]. Using diameter, strength of the electric field about 15 kV/cm and the current density 1.6 10$^7$ A/cm$^2$ were estimated at, e.g., HRS to LRS transition.

Figure 2 shows *dV/dI* and *I–V* (in inset) characteristics of PC with WTe$_2$ at room temperature, where difference in the resistance between LRS and HRS reaches more than two orders of magnitude (see also Fig. S8 in Supplement, where similar result is shown for the case of MoTe$_2$). There is instability in *I-V* at LRS to HRS transition above 150 mV. At the same, time analogous transition at helium temperature is quite sharp (see, e.g., Fig. 1). This is apparently due to enhanced fluctuation at structural transition at higher temperature. Figure 2 demonstrates also comparison of *dV/dI* for the same PC measured with time interval of about 12 hours to show their reproducibility (or stability of switching effect under ambient conditions).

Similar switching effects were observed also for TaRhTe$_4$ and TaRuTe$_4$ (see Figs. S1 and S2 in Supplement). "Hard" PCs, where instead of silver paint a thin Ag wire was used, also demonstrate switching effect (see Figs. S2 and S3 in Supplement). A disadvantage of "hard" PCs is their weak mechanical stability, what makes difficult to keep them stable for a relative long time and especially by varying a temperature, therefore most measurements were carried out on "soft" PCs.

Figure 3 shows distribution of the switching voltages for LRS to HRS and HRS to LRS transitions for different PCs and for all compounds at helium and room temperatures. Here, the most data are presented for MoTe$_2$ and TaIrTe$_4$ compounds (6 PCs for each compound). First of all, it is seen, that polarity of each transition (LRS to HRS or HRS to LRS) can be both positive and negative for different PCs with almost equal probability. For example, 3 PCs with MoTe$_2$ or

TaIrTe$_4$ have LRS to HRS transition at positive polarity and 3 PCs vice versa at negative. However, if the switching occurs at a given polarity then it is absent at the opposite polarity, at least in the same voltage range. Perhaps the atomic composition of the termination layer or direction of the ferroelectric moment play a role here. This point requires further investigation. Also, switching voltage has some distribution but is, on the average, smaller at a room temperature.

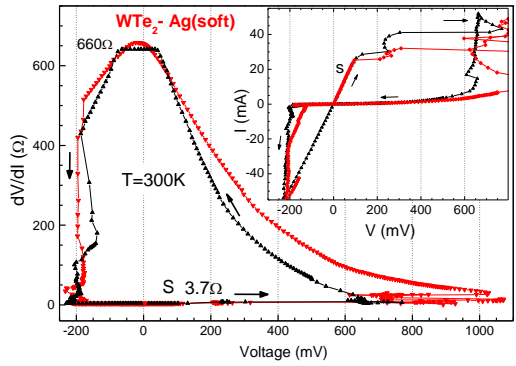 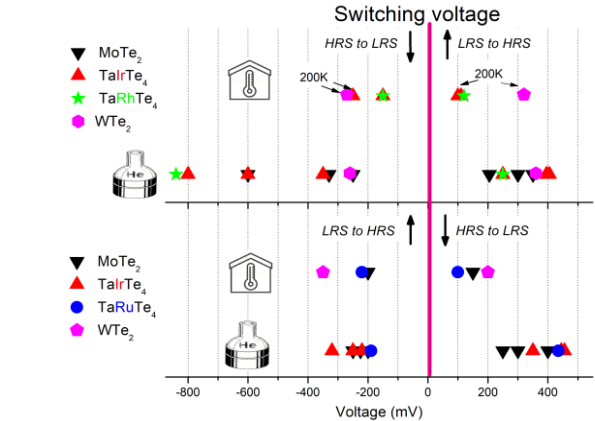

**Figure 2.** Differential resistance *dV/dI* and *I-V* curves (inset) behavior for "soft" contact based on WTe$_2$ at room temperature. Two *dV/dI* of "soft" contact WTe$_2$–Ag measured with time interval of 12 hours after storage this contact under ambient condition. "s" marks the start of measurements and arrows show direction of the current scan. Note, the zero bias resistance is changed more than two order of magnitude. Inset shows *I–V* curve corresponding to *dV/dI* from the main panel.

**Figure 3.** Distribution of the switching voltages for different PCs with studied compounds at helium (dewar) and room (box) temperatures. Upper (bottom) panel shows PCs where switching from LRS (HRS) to HRS (LRS) take place at positive (negative) voltage. Switching voltages for LRS to HRS transition are marked by up arrows and for HRS to LRS transition are marked by down arrows. For two PCs, switching voltages are shown at 200K (tilted arrows).

Finally, Figure S4 (in Supplement) demonstrates the detailed temperature dependence of *dV/dI*, where switching effect was found by us for the first time at room temperature. There, the lack of switching at a low temperature was connected with necessity to apply a larger bias, while our setup has maximal current output of about 50 mA, therefore the maximal bias was limited by the low PC resistance. Detailed measurement of zero-bias resistance of this PC at different temperatures allowed us to estimate their size and, respectively, the electric field strength and current density. As a result, the estimated strength of an electric field at switching bias was about 2.5 times lower for this PC as compared to PC at helium temperature (see Fig. 1(d)). It should be noted here, that the estimation of the electric field gives it a lower value, since we assume the formation of single metallic contact or conducting channel. In the case of formation of several PCs in parallel, their size will be smaller and the electric field will be correspondingly larger.

Note that *dV/dI* of MoTe$_2$ PC in LRS (see, e.g., black curve in Fig.1(c), (d)) has "metallic" behavior at low voltages, which transforms to the "semiconducting-like" decrease of *dV/dI* at higher bias (see also Fig. S4(a), (b) for more details) producing a distinct maximum in *dV/dI* around 200 mV similar to that what we observed in the case of WTe$_2$ [5, 9]. Interesting, that metallic to semiconducting like transition was observed around 200 K in the temperature dependence of resistance of 6 nm thick MoTe$_2$ sample (see Fig. 4S in Supplement of Ref. [22]). It seems that the reducing of the dimension promotes the emergence of a "semiconductor-like" behavior in MoTe$_2$.

**Discussion**

A switching effect in TMT was reported by Zhang *et al.* [13] in "vertical devices made of MoTe$_2$ and Mo$_{1-x}$W$_x$Te$_2$ (x<0.1) layers" (from 6 nm to 36 nm thick) sandwiched between metallic electrodes. Interestingly that the "active device area" (in average 500x400 nm$^2$) in [13] is close to the estimated size of PCs of several hundred nm (see Fig. 1(d) and S4(c)). Zhang *et al.* [13] explained the switching assuming a formation of a conductive filament with a transient structure (named 2H$_d$) by applying an external electrical field. The phase 2H$_d$, they considered as a distorted 2H phase, which exhibits electrical properties that range from semiconducting to metallic. Zhang *et al.* [13] believe that MoTe$_2$ (or Mo$_{1-x}$W$_x$Te$_2$) undergo a reversible structural transition from a semiconducting 2H to a high (or more) conductive 2H$_d$ state. Datye *et al.* [14] reported the observation of similar switching in structure contained layered MoTe$_2$ (10-55 nm thick) sandwiched between Au electrodes. They used scanning thermal microscopy to monitor temperature of the top Au electrode and observed hot spot of about 1 μm size. They concluded that the switching "is likely caused by the breaking and forming of Au conductive plugs between the electrodes", when significant Joule heating promotes Au migration.

We used "bulk" samples with thickness of about 100 μm to create PCs. In our case the observed "bipolar" switching from LRS to HRS occurs only at definite polarity above a definite threshold. It's important, that we started each measurement from the "metallic" low resistance state (market by "s" in figures) and we observed transition to the high resistive "semiconducting" type state if the applied voltage is above some threshold. It excludes Ag-filaments as a driving force for the switch effect. We have also tried to start from the semiconducting 2H-phase of MoTe$_2$, but we observed no switching effect. This behavior and the fact that observed "bipolar" switching from LRS to HRS (and opposite from HRS to LRS) occurs only at definite polarity unambiguously indicates a nonthermal mechanism. Therefore, our effect is different from resistive switching described by Zhang *et al.* [13] and Datye *et al.* [14]. Since thermal effects are disregarded, the only effect of electric field and/or high current density in our PCs must play a role.

Further, high (or more) conductive state (named here as LRS) demonstrates metallic behavior, as Fig. 1(d) and Figs. S3, S4(a), (b) show, where, at first, *dV/dI* increases with voltages at zero-bias. Hence, in our case, LRS corresponds to semimetallic 1T`or T$_d$ phase and low conductive state (or HRS) with sharp zero-bias maximum in *dV/dI* is formed by applying a voltage. It can be due to some formation of a distorted state (or charge density waves) or due to enhanced resistance at the domain wall between two domain with opposite polarity which were discussed in [15, 16, 24] with respect to MoTe$_2$ and WTe$_2$. The domain formation in PC is likely caused by high electric field in its core (see Fig. 4(a)). That is quite possible, given the fact that according to the recent observation, native metallicity and ferroelectricity can coexist in MoTe$_2$ and WTe$_2$ [15,16, 24]. Yuan *et al*. [15] relate the origin of ferroelectricity in the MoTe$_2$ monolayer to spontaneous symmetry breaking due to relative displacements of Mo atoms and Te atoms as a result of the formation of a distorted d1T phase. Further, Fei *et al.* [16] observed that two- or three-layer WTe$_2$ exhibits spontaneous out-of-plane electric polarization that can be switched using gate electrodes and the polarization states can be differentiated by their conductivity. Moreover, Sharma *et al.* [24] provide evidence that native metallicity and ferroelectricity coexist also in bulk crystalline WTe$_2$. They have also demonstrated that WTe$_2$ has switchable spontaneous polarization and natural ferroelectric domains with a distorted circular profile with an average domain size of ~20 to 50 nm. That is, ferroelectricity is a bulk property of WTe$_2$ and is not limited to few-layer samples only. Therefore, we assume that when relative atomic displacements of Mo atoms and Te atoms produces polarization (in other words electric field), then inverse effect must also take place. Namely, high electric field in PC can induce mentioned atomic displacement that leads to the phase transition and/or appearance of domain in PC (see Fig. 4(a)) with different properties. In this case, the new phase (something like distorted d1T) must manifest semiconducting like behavior. Then, observed switching in the formed in PC core domain can be due to reversible phase transition between distorted/intermediate (maybe d1T) and semimetallic T$_d$ or 1T` phase triggered by high electric field.

Considering that the title materials belong to the family of Weyl semimetals with topologically protected surface states, the discovery and exploitation of specific properties of these surface states in homo/hetero-structures is of a great challenge. Note that the investigated materials have the same T$_d$ crystal structure at low temperatures, which has broken inversion symmetry. It results in an uncompensated dipole, resulting in polarization along the c-axis, which can exist in the T$_d$↑ and T$_d$↓ states due to the asymmetric Te bonding environments [25]. Imagine, that under the action of high electric field a domain (see Fig. 4(a)) has a certain polarization, let's say T$_d$↑. Therefore, a high electric field of opposite polarity can flip the dipole moment of domain with the corresponding change of the crystal structure from T$_d$↑ to T$_d$↓. The interface between T$_d$↑ and T$_d$↓ is T` phase (see Fig. 4(b)). Thus, the electron scattering on the domain wall between T$_d$↑ and T$_d$↓ phase with specific surface states can lead to the increased resistance and simulate HRS behavior. Of course, this model requires sophisticated theoretical calculations, that have not yet been completed. It should be recalled here that the *dI/dV* at HRS display a linear behavior (see inset of Fig. 1c and Figs. S5, S6, S7 in Supplement). As can also be seen on Figures in Supplement, that the *dV/dI* has a logarithmic behavior for a certain bias range. This observation is interesting and it must be considered when developing a theory of switching effect.

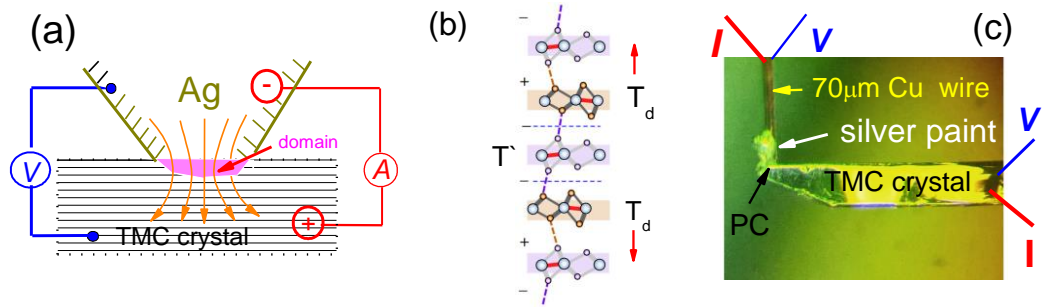

**Figure 4.** Representation of point contact. (a) Model of PC between TMT crystal and Ag-tip showing current and voltage leads and "domain", where an electric field is maximal. Arrows show spreading of a current in PC. (b) schematic model of domain wall between $T_d\uparrow$ and $T_d\downarrow$ states containing one 1T´ unit as bridge according to [25]. (c) Photo image of "soft" PC, where silver paint touches the sample and connects to a thin Cu wire Ø70µm. Cu wire and sample are connected to an electrical circuit by *I* and *V* leads.

**Conclusion and outlook**

We report a new resistive switching in PCs based on series of TMT compounds such as $Me\text{Te}_2$ (*Me*=Mo, W) and Ta$Me$Te$_4$ (*Me*= Ru, Rh, Ir), what is summarized in Fig.3. The switching effect consists of, up to two orders of magnitude change in the PC resistance, which increases with lowering of temperature. The origin of the effect is likely due to reversible modification of the crystal/electronic structure of TMT and creation of switchable domain in PC core under the application of high electric field. As a result, the transition takes place between the low resistive metallic and the high resistive semiconducting-type states depending on domain structure. Besides, presence of domain walls with topological interfacial states can also play a role. Undoubtedly, used by us a simple method of device preparation has a great advantage in finding suitable materials and can be used, at least, to look for resistive switching effect before functionalizing them in nano-electronic application. It worth noting here, that the first observation of the resistive switching effect, e. g., in strongly correlated high-$T_c$ materials [26] and manganese-based perovskite oxides with collosal magnetoresistance [27], was also observed by using of PC technique, demonstrating its capabilities. Summing up, our observation of the resistive switching effect in series of TMT compounds expands significantly the range of materials promising for RRAM development [13, 28], for ferroelectric phase change transistors [29], for other nano-electronic applications [30] and opens a straight path for discovering the effect in other metallic layered materials, what paves the way to electronics with new improbable functionalities. On the other hand, the understanding underlying physics of the switching processes is of great importance, which can open the avenue to application of topological surface states.

**Experimental Section**

*Synthesis.* Bulk single crystals of $MoTe_2$, $WTe_2$ and other TMT compounds were grown with Te flux. To avoid contamination, the mixing and weighting were carried out in an Ar-filled glove box. Amounts of 0.5 g of Mo, W powder and 10 g Te were mixed and placed in an evacuated quartz ampoule. The ampoule was placed in a box furnace and slowly heated to 1000 °C and cooled down slowly to 800 °C followed by a hot centrifuge to remove the excess Te-flux. Single crystals were grown having a needle-like shape with a layered morphology. The as-grown crystals were characterized by SEM in EDX mode for compositional analysis and by x-ray diffraction for structural analysis. More details of the crystal growth and characterization are reported in Refs. [31,32,33].

*Point contact spectroscopy.* PCs were prepared by touching of a thin Ag wire to a cleaved at room temperature flat surface of TMT needle-like single-crystal flake or contacting its edge by this wire. The estimating diameter of the point contact is of the order of 100 -300 nm. So-called "soft" PCs made by dripping of a small drop of silver paint onto the cleaved TMT surface/edge (see Fig. 4(c)). The latter type of PCs demonstrates better stability versus temperature change. Thus, we conducted resistive measurements on the hetero-contacts between a normal metal (Ag or silver paint) and the TMT. We measured current–voltage characteristics *I–V* of PCs and their first derivatives *dV/dI(V)*. The first derivative or differential resistance $dV/dI(V) \equiv R(V)$ was recorded by scanning the dc current *I* on which a small ac current *i* was superimposed using a standard lock-in technique. The measurements were performed in the temperature range from liquid helium up to the room temperature.

**Supporting Information**
Supporting Information is available from the Wiley Online Library or from the author.

**Acknowledgments**
We thank A. V. Terekhov and S. Gaβ for the technical assistance. This work was financially supported by the Volkswagen Foundation in the frame of Trilateral Initiative. YGN, DLB and OEK are grateful for support by the National Academy of Sciences of the Ukraine under project Φ4-19 and would like to thank the IFW Dresden for hospitality. Support by Deutsche

Forschungsgemeinschaft (DFG) through Grant No: AS 523/4-1 is acknowledged. SA, BB, DE also acknowledge support of DFG through Projekt No:405940956.

**Conflict of Interest**

The authors declare no conflict of interest

**References**

[1] S. Manzeli, D. Ovchinnikov, D. Pasquier, O. V. Yazyev and A. Kis, *Nat. Rev. Mater.* **2017,** *2,* 17033.

[2] B. Yan, C. Felser, *Annu. Rev. Condens. Matter Phys*. **2017**, *8*, 337.

[3] Q. H. Wang, K. Kalantar-Zadeh, A. Kis, J. N. Coleman, M. S. Strano, *Nature Nanotechnol*. **2012,** 7, 699.

[4] Mazhar N. Ali, Jun Xiong, Steven Flynn, Jing Tao, Quinn D. Gibson, Leslie M. Schoop, Tian Liang, Neel Haldolaarachchige, Max Hirschberger, N. P. Ong & R. J. Cava, *Nature* **2014**, *514*, 205.

[5] S. Thirupathaiah, Rajveer Jha, Banabir Pal, J. S. Matias, P. Kumar Das, P. K. Sivakumar, I. Vobornik, N. C. Plumb, M. Shi, R. A. Ribeiro, and D. D. Sarma, *Phys. Rev. B* **2017**, *95*, 241105(R).

[6] Seunghyun Khim, Klaus Koepernik, Dmitry V. Efremov, J. Klotz, T. Förster, J. Wosnitza, Mihai I. Sturza, Sabine Wurmehl, Christian Hess, Jeroen van den Brink, and Bernd Büchner, *Phys. Rev. B,* **2016**, *94*, 165145.

[7] Yanpeng Qi, Pavel G. Naumov, Mazhar N. Ali, Catherine R. Rajamathi, Walter Schnelle, Oleg Barkalov, Michael Hanfland, Shu-Chun Wu, Chandra Shekhar, Yan Sun, Vicky Süß, Marcus Schmidt, Ulrich Schwarz, Eckhard Pippel, Peter Werner, Reinald Hillebrand, Tobias Förster, Erik Kampert, Stuart Parkin, R.J. Cava, Claudia Felser, Binghai Yan & Sergey A. Medvedev, *Nat. Commun.* **2016**, *7*, 11038.

[8] Yu. G. Naidyuk, O. E. Kvitnitskaya, D. L. Bashlakov, S. Aswartham, I. V. Morozov, I. O. Chernyavskii, G. Fuchs, S.-L. Drechsler, R. Hühne, K. Nielsch, B. Büchner and D. V. Efremov, *2D Mater*. **2018,** *5*, 045014.

[9] Yu. G. Naidyuk, D. L. Bashlakov, O. E. Kvitnitskaya, S. Aswartham, I. V. Morozov, I. O. Chernyavskii, G. Shipunov, G. Fuchs, S.-L. Drechsler, R. Hühne, K. Nielsch, B. Büchner, D. V. Efremov, *2D Mater*. **2019**, *6,* 045012.

[10] F. Tang, H. C. Po, A. Vishwanath, and X. Wan, *Nat. Phys.* **2019**, *15*, 470.

[11] M. Ezawa, Sci. Rep. **2019**, *9*, 5286.

[12] Wang, Z., Wieder, B. J., Li, J., Yan, B. & Bernevig, B. A. **Phys. Rev. Lett**. **2019**, 123, 186401.

[13] F. Zhang, H.R. Zhang, S. Krylyuk, C. A. Milligan, Y. Q. Zhu, D. Y. Zemlyanov, L. A. Bendersky, B. P. Burton, A. V. Davydov, J. Appenzeller, *Nat. Mater.* **2019***, 18,* 55.

[14] I. Datye, M. M. Rojo, E. Yalon, S. Deshmukh, M. J. Mleczko and E. Pop, *NanoLett.* **2020,** *20***,** 1461.

[15] Z. Fei, W. Zhao, T. A. Palomaki, B. Sun, M. K. Miller, Z. Zhao, J. Yan, X. Xu and D. H. Cobden, *Nature,* **2018,** *560,* 336.

[16] S. Yuan, X. Luo, H. L. Chan, C. Xiao, Y. Dai, M. Xie, and J. Hao, *Nat.Commun.* **2019***,* 10, 1775.

[17] N. A. Benedek and T. Birol, J. Mater. Chem. C **2016**, *4*, 4000.

[18] Yu. G. Naidyuk, I. K. Yanson, Point-Contact Spectroscopy, *Springer Series in Solid-State Sciences.* **2005**, vol 145 (New York: Springer).

[19] B. I. Verkin, I. K. Yanson, I. O. Kulik, O. I. Shklyarevski, A. A. Lysykh and Yu. G. Naydyuk, *Solid State Communs.* **1979,** 30, 215.

[20] D. Rhodes *et al*., arXiv preprint arXiv:1905.06508

[21] R. Clarke, E. Marsegilia and H. P. Hughes, *Philos. Mag.* **1978**, 38, 121.

[22] C. Cao, X. Liu, X. Ren, X. Zeng, K. Zhang, D. Sun, S. Zhou, Y. Wu, Y. Li, and J.-H. Chen, *2D Mater.* **2018,** *5***,** 044003.

[23] S. Song, D. H. Keum, S. Cho, D. Perello, Y. Kim and Y. H. Lee, *Nano Lett.* **2016**, *16*, 188.

[24] P. Sharma, F.-X. Xiang, D.-F. Shao, D. Zhang, E. Y. Tsymbal, A. R. Hamilton, J. Seidel, *Sci. Adv*. **2019**, *5*, eaax5080.

[25] F.-T. Huang, S. J. Lim, S. Singh, J. Kim, L. Zhang, J.-W. Kim, M.-W. Chu, K. M. Rabe, D. Vanderbilt and S.-W. Cheong, *Nat. Commun*. **2019**, *10,* 4211.

[26] L. F. Rybaltchenko, N. L. Bobrov, V. V. Fisun, I. K. Yanson, A. G. M. Jansen and P. Wyder, *European Physical Journal B,* **1999,** *10*, 475.

[27] M. A. Belogolovskii, Yu. F. Revenko, A. Yu. Gerasimenko, V. M. Svistunov, E. Hatta, G. Plitnik, V. E. Shaternik and E. M. Rudenko, *Low Temp. Phys.* **2002,** 28, 391.

[28] Q. Zhao, Z. Xie, Y.-P. Peng, K. Wang, H. Wang, X. Li, H. Wang, J. Chen*,* H. Zhang, X. Yan, *Mater. Horiz.* **2020,***7***,** 1495.

[29] W. Hou, A. Azizimanesh, A. Sewaket, T. Peca, C. Watson, M. Liu, H. Askari and S. M. Wu, *Nat. Nanotech.* **2019**, 14, 668.

[30] J. Su, K. Liu, F. Wang, B. Jin, Y. Guo, G. Liu, H. Li and T. Zhai, *Adv. Mater. Interfaces,* **2019,** *6*, 1900741.

[31] A.-S. Pawlik, S. Aswartham, I. Morozov, M. Knupfer, B. Büchner, D. V. Efremov and A. Koitzsch, *Phys. Rev. Materials,* **2018,** *2*, 104004.

[32] S. Khim, K. Koepernik, D. V. Efremov, J. Klotz, T. Förster, J. Wosnitza, M. I. Sturza, S. Wurmehl, C. Hess, J. van den Brink and B. Büchner, *Phys. Rev. B* **2016***, 94,* 165145.

[33] G. Shipunov, D. V. Efremov, B. Büchner, S. Aswartham, **2020** (in preparation).

# Supporting Information

to the paper by Naidyuk et al.: "**Switchable domains in point contacts based on transition metal tellurides**"

1) Examples of resistive switching in PCs with TaRhTe$_4$ and TaRuTe$_4$ compounds.

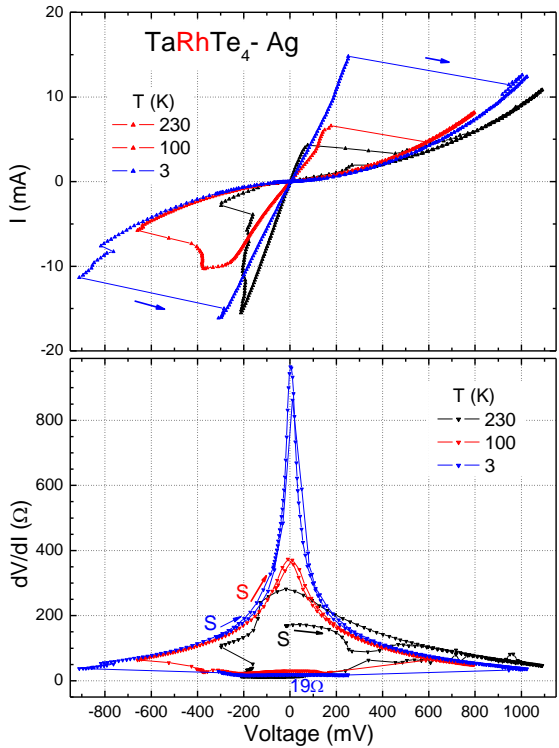 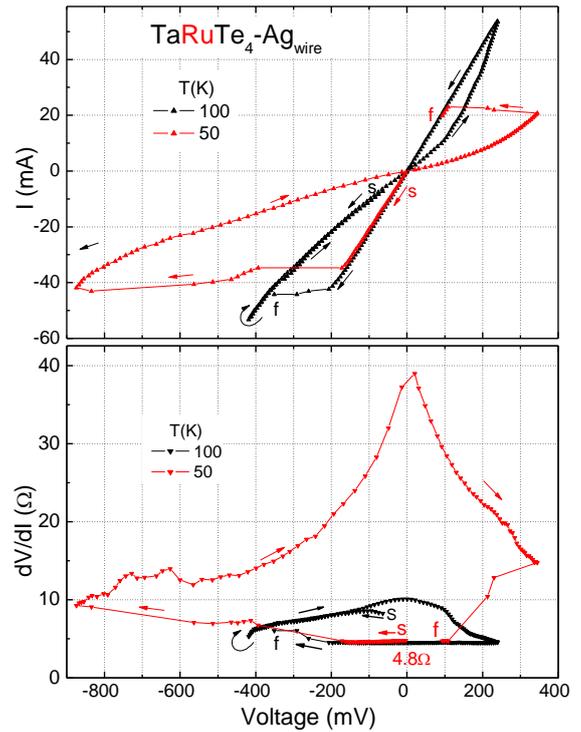

**Figure S1**. Upper panel: *I–V* curves of "soft" PC TaRhTe$_4$–Ag measured at different temperatures. Bottom panel: *dV/dI* of the same "soft" PC."s" marks the start of measurements from HRS. PC resistance is 19 Ohm in LRS at 3K.

**Figure S2**. Upper panel: *I–V* curves of "hard" PC TaRuTe$_4$–Ag measured at two temperatures. Bottom panel: *dV/dI* of the same "hard" PC."s" marks the start of measurement and "f" is the finish, arrows show direction of the current sweep. PC resistance is 4.8 Ohm in LRS at 50K.

2) Examples of resistive switching in "hard" PCs and in low-ohmic "soft" PC.

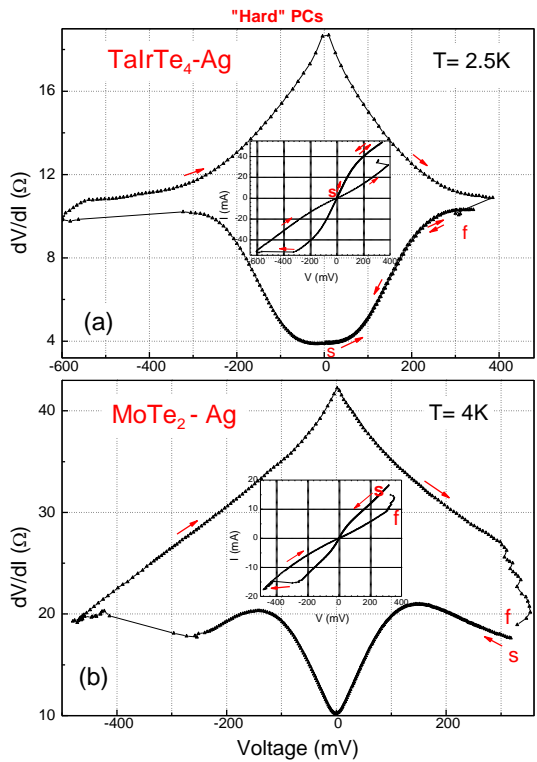

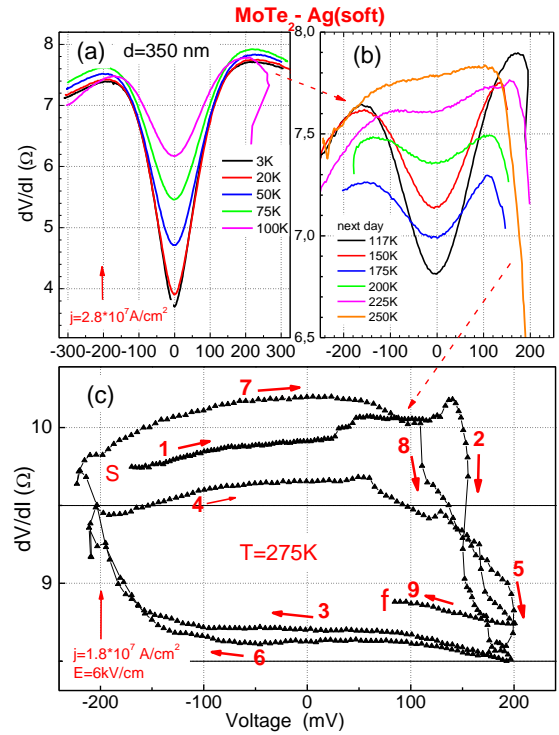

**Figure S3**. *dV/dI* of two "hard" PCs made by mechanical contact of thin 70 μm Ag wire to TaIrTe$_4$ and MoTe$_2$ crystals at helium temperature. Insets show *I–V* characteristics for same PCs from the main panels.

**Figure S4.** (a), (b) Evolution of *dV/dI* of MoTe$_2$ "soft" contact with temperature. Here a sharp decline of *dV/dI* at a positive voltage may indicate the start of switching. (c) *dV/dI* of the same contact with switching effect successively measured "clock-wise" repeatedly at 275K. The diameter of this contact was estimated as 350 nm using the temperature dependence of their zero bias resistance. Strength of the electric field and the current density were estimated at LRS to HRS transition as shown by vertical arrows.

3) Details of *dV/dI* and *dI/dV* behavior in PCs.

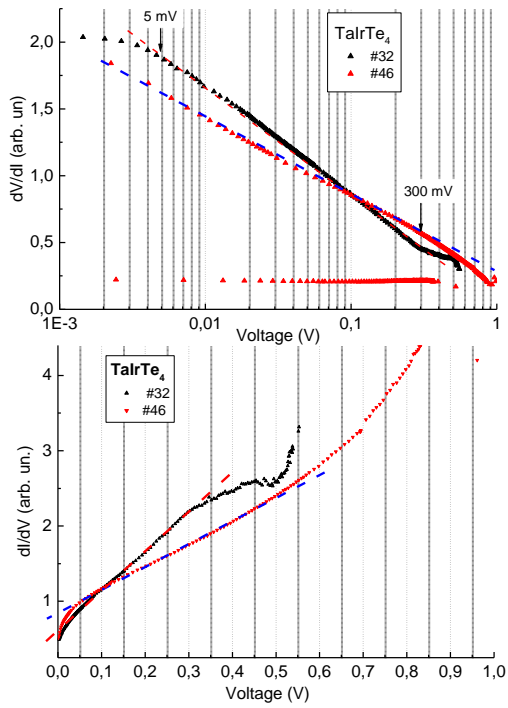

**Figure S5**. Upper panel: *dV/dI* curves of two "soft" PCs TaIrTe$_4$–Ag from Fig. 1(a) in log-scale. Bottom panel: *dI/dV* for the same PCs. The dash lines in the figures are guide for the eye.

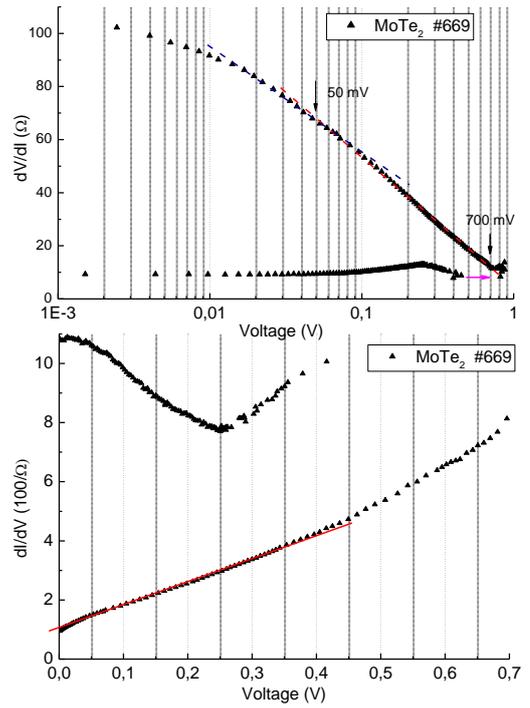

**Figure S6.** Upper panel: *dV/dI* curves of "soft" PC MoTe$_2$–Ag from Fig 2 in log-scale. Bottom panel: *dI/dV* for the same PC. The dash lines in the figures are guide for the eye.

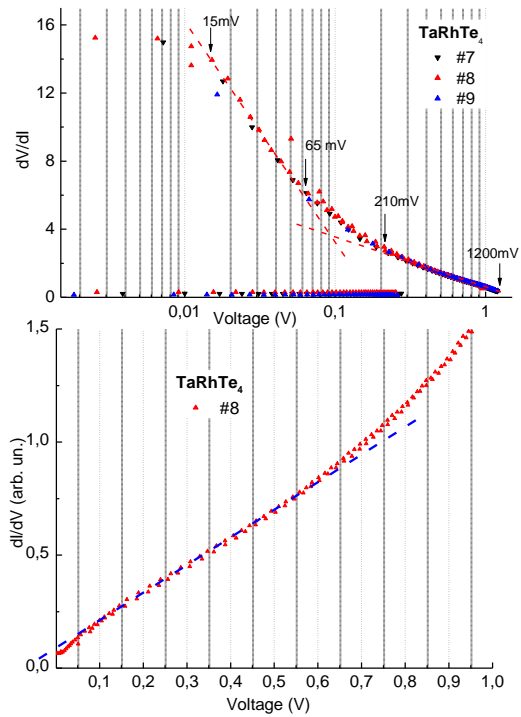

**Figure S7**. Upper panel: *dV/dI* curves of "soft" PC TaRhTe$_4$–Ag for three successive sweeps (different symbols) in log-scale. Bottom panel: *dI/dV* for the same PC. The dash lines in the figures are guide for the eye.

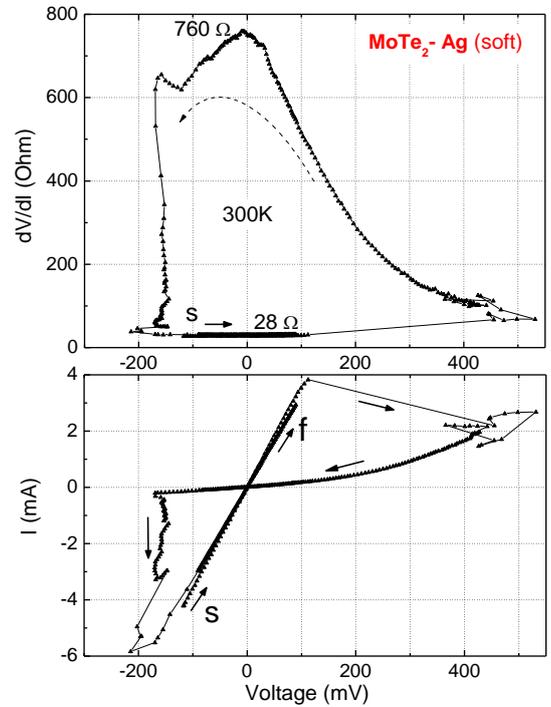

**Figure S8**. Differential resistance *dV/dI* and *I-V* curves behavior for soft contact based on MoTe$_2$ at room temperature.